\documentstyle[twocolumn,aps,pra,epsf]{revtex}
\newcommand{\be}{\begin{equation}}
\newcommand{\ee}{\end{equation}}
\newcommand{\bea}{\begin{eqnarray}}
\newcommand{\eea}{\end{eqnarray}}
\def\n{\noindent}
\def\etal{{\sl et al.}}

\begin{document}

\title{Interaction-free measurement and forward scattering}
\author{Tam\'as Geszti}
\address{Department of the Physics of Complex Systems, \\
E\"otv\"os University; H-1088 Budapest, Hungary \\
{e-mail: \tt geszti@hercules.elte.hu}
}
\maketitle

\begin{abstract}
Interaction-free measurement is shown to arise from the forward-scattered 
wave accompanying absorption: a ``quantum silhouette" of the absorber.
Accordingly, the process is not free of interaction. For a perfect absorber 
the forward-scattered wave is locked both in amplitude and in phase. 
For an imperfect one it has a nontrivial phase of dynamical origin 
(``colored silhouette"), measurable by interferometry. Other examples 
of quantum silhouettes, all controlled by unitarity, are briefly discussed.
\end{abstract}


Interaction-free Measurement (IFM) is a term first used by 
Renninger\cite{ren}, then by Dicke\cite{di} to label the paradoxical 
observation that a negative-result quantum measurement, apparently without 
interaction, modifies the wave function of the non-detected object. Dicke's 
resolution of the paradox is essentially to point out that the interaction 
involved in the act of measurement creates an entanglement between probe 
and object, thereby changing the state of their assembly irrespective of the 
outcome of the measurement, which is therefore not interaction-free. The 
present paper lends more support to the latter statement, based on a slightly 
different argument. Our aim is to present a clear physical picture of 
IFM, rather than suggesting strategies for improved performances.

The idea of IFM gained popularity with the paper of Elitzur and 
Vaidman\cite{ev} who amplified the argument by inventing an 
efficient interferometric setting for it. They suggested to place a perfect 
absorber in one arm of a Mach-Zehnder photon interferometer (Fig. 1). 
Without the absorber, on one of the exit ports (the ``dark port", in the 
figure: $D_a$) there is no output, because of destructive interference. 
Inserting the absorber changes the interference pattern; in particular, it 
suspends the destructive interference. Then some photons reach the previously 
silent detector, indicating that an absorber is there. That kind of 
interferometric detection of an absorber, by observing a photon that 
avoids absorption, is called interaction-free measurement in the growing 
recent literature on the subject.

\begin{figure}
\epsfxsize=3.0in \epsfysize=1.4in \epsffile{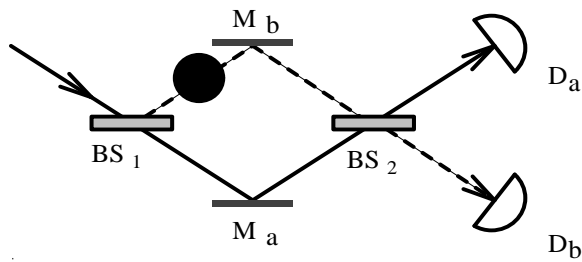} \vskip 0.2cm
\caption{Interaction-free measurement in a Mach-Zehnder interferometer. 
The solid circle is the ``bomb" (see text). Mode  $a$ is denoted by a solid 
line, mode $b$ by a dashed line, $D_a$ and $D_b$ are the respective 
detectors, $BS_1$ and $BS_2$ the two 50\% beamsplitters, $M_a$ and 
$M_b$ are mirrors.}
\end{figure}

Elitzur and Vaidman\cite{ev} dramatize the situation by calling the absorber 
a ``bomb" that detects the act of absorption by exploding with 100\% 
efficiency. The starting point of our discussion is this: the bomb plays a 
double role in the scheme; it is the combination of an absorber and a 
detector of its excited state. The act of its explosion - a quantum 
measurement process - should be distinguished from the preceding microscopic, 
unitary process of absorbing the photon. Detectors and detonators work on the 
final state of the unitary process, and they work under the constraint 
of wave-particle duality\footnote{That connection with wave-particle duality 
has been pointed out in Ref. \cite{kwhz}. This is a strange case however: a 
choice between final states and detectors of different nature.}: only one term 
of the branching superposition can be actually detected. The absorptive branch 
is just one of the possible outcomes. However, even if not all terms of 
the superposition get detected, they are all there, leaving their 
imprints -- silhouettes -- on each other, forced by unitarity.

In what follows, we first describe that simplest situation along the line 
introduced by Elitzur and Vaidman\cite{ev}\footnote{An analysis from the 
point of view of photon - absorber entanglement, somewhat similar to ours, 
has been presented by Kwiat \etal\cite{kwz}}. We propose to use the language 
of scattering theory, in the sense of locating causal changes of the quantum 
state due to the presence of the absorber, as compared to its absence. In 
that context, the notion of forward scattering -- an essential aspect of 
inelastic scattering, including absorption, as succinctly expressed by the 
Optical Theorem\cite{sch,ll,ct} -- offers a natural framework to think 
about the phenomenon of IFM with all its 
recent versions\cite{kwhzk,kss,pav,pp,kbf,hs}\footnote{Forward scattering in 
the context of IFM has been mentioned by Krenn \etal\cite{kss} as a side 
effect. The viewpoint taken in the present note is that it is the main 
thing about IFM.}. 

``Forward scattering" is just another name for the wave-optical shadow cast 
by a scattering and eventually absorbing object. The image of the object 
formed by shadow scattering, visible against the interferometrically set 
featureless background, is what we propose to call a ``quantum silhouette", 
as discussed in more detail below. 

The forthcoming analysis intends to describe what happens to a flying 
one-photon wave packet entering the interferometer. As usual in scattering 
theory, as a limiting case of long wave packets, one can think 
of quasi-one-dimensional plane waves; in that limiting case, however, care 
must be taken that the scattered wave is outgoing from the scatterer. 
The transverse extension of the quasi-plane wave is thought to be large 
with respect to the wavelength and small with respect to the size of 
the macroscopic absorber; in technical terms, one may think of a Gaussian 
or Bessel beam\cite{st}. 

The Mach-Zehnder interferometer supports two photon modes \(a\) and 
\(b\), shaped by the respective mirrors indicated in Figure 1. Each of 
the beamsplitters causes \(a\leftrightarrow b\) transitions, for the 50\% case 
with amplitude \(i/\sqrt2\), the amplitude of transmission without changing 
the mode being \(1/\sqrt2\).

 Interaction of the photon with the absorber changes the state of the latter 
as well. Thereby the absorber gets entangled with the photon field. For 
simplicity, we want to take into account only two states of the absorber: 
ground state \(G\) and excited state \(E\).

Each mode can be populated by a number of photons. The events we want 
to describe include one-photon propagation and resonant absorption, therefore 
they take place on a basis of three vectors: one photon on mode \(a\) or 
\(b\) with the absorber in ground state, or zero photon with the absorber in 
excited state. Accordingly, we use a three-line column vector notation for 
quantum states, with the basis

\be
|1_a0_b~G\rangle = \pmatrix{1\cr 0\cr 0\cr};~~
|0_a1_b~G\rangle = \pmatrix{0\cr 1\cr 0\cr};~~
|0_a0_b~E\rangle = \pmatrix{0\cr 0\cr 1\cr}.
\label{basis}
\ee

\n On this basis the beamsplitter and the absorber are represented by the 
respective unitary matrices

\be
\bf{U_{BS}}=
\pmatrix{1/\sqrt2&i/\sqrt2&0\cr i/\sqrt2&1/\sqrt2&0\cr 0&0&1};~~~
\bf{U_{AB}}=\pmatrix{1&0&~~0\cr 0&0&-1\cr 0&1&~~0\cr},
\label{unitaries}
\ee

\n the latter describing a local interaction between the perfect absorber on
mode \(b\) with the photon field. Of course, in accordance with unitarity, 
the probability amplitude lost by photon absorption reappears transferred 
to the 0-photon, excited absorber state.

The whole sequence of unitary evolution is represented as follows:

\be
\pmatrix{1\cr 0\cr 0\cr} 
    \rightarrow\pmatrix{1/\sqrt2\cr i/\sqrt2\cr 0\cr}
    \Rightarrow\pmatrix{1/\sqrt2\cr 0\cr i/\sqrt2\cr}
    \rightarrow\pmatrix{{\bf1/2}\cr {\bf i/2}\cr i/\sqrt2}.
\label{bomb}
\ee

\n The plain arrows correspond to the two beamsplitters; the central, double 
arrow represents the absorber (other parts of the bomb being not involved 
yet).

We assume three ideal detectors, each covering one component on the 
above basis. Then the sequence of unitary events is concluded by 
a quantum measurement stage in which the system is forced to randomly 
choose one of the respective mutually exclusive possibilities: to detect a 
photon on mode \(a\) (detector \(D_a\)), a photon on mode \(b\) 
(detector \(D_b\)), or the excited state of the absorber (its macroscopic 
detector being the famous bomb or a more friendly laboratory version of it). 

Let us turn to the scattering theory language. If we want to locate causal 
changes caused by the absorber, we have to define a reference sequence of 
unitary events -- the  ``incoming wave" -- with no absorber ({\sl i.e.} 
\(\bf{U_{AB}}\) replaced by unity):

\be
\pmatrix{1\cr 0\cr 0\cr} 
	\rightarrow\pmatrix{1/\sqrt2\cr i/\sqrt2\cr 0\cr}
	\Rightarrow\pmatrix{1/\sqrt2\cr i/\sqrt2\cr 0\cr}
	\rightarrow\pmatrix{0\cr i\cr 0\cr}.
\label{nobomb}
\ee

\n As expected, there is full destructive interference on mode \(a\). 

The difference between Equations (\ref{bomb}) and (\ref{nobomb}) is the 
``scattering amplitude"

\be
\pmatrix{0\cr 0\cr 0\cr} 
    \rightarrow\pmatrix{0\cr 0\cr 0\cr}
    \Rightarrow\pmatrix{0\cr -i/\sqrt2\cr i/\sqrt2\cr}
    \rightarrow\pmatrix{1/2\cr -i/2\cr i/\sqrt2\cr}.
\label{scattered}
\ee

\n More specifically: the third line of the final column vector is the 
amplitude of photon absorption (as part of the microscopic, unitary 
evolution), not yet detected, but detectable by the bomb if the subsequent 
measurement process chooses that branch of the superposition, whereas 
the first two lines represent the ``forward-scattered wave" added to the 
one-photon components by the absorber: unitary imprints of the same 
interaction process, carrying information about it, much the way a person 
can be recognized from her or his silhouette. 

What one usually calls IFM is the non-zero output on the first line of the 
final column vector of Equation (\ref{bomb}) or (\ref{scattered}): neglecting 
detector background noise, each non-absorbed photon counted by detector 
\(D_a\) indicates the presence of an absorber. However, the same 
non-absorbed photons inform about the same absorber also by reducing the 
number of counts on \(D_b\) by a factor of 4, which is therefore an 
important part of the scenario. That reduction, apparent in Equation 
(\ref{bomb}), comes about by destructive interference between the incoming 
wave (\ref{nobomb}) and the forward-scattered one (\ref{scattered}). The 
same feature is present in the case of an imperfect absorber, see below.

The above discussion explains why we think that ``interaction-free" is a 
misleading adjective, unless we use it just as a label. As long as we follow 
the initial, microscopic stage, the evolution is unitary, and the appearance of 
a non-zero probability amplitude on the absorptive state must be accompanied 
by a corresponding change somewhere on the non-absorptive components, 
everything being controlled by the same interaction.\footnote{The situation 
is somewhat similar, although not identical, to the so-called no-cloning 
theorem of quantum computation. There, one encounters the impossibility of 
having a unitary projector (apart from unity). The present situation is rather 
the impossibility for a quantum thief to take his pickings away without leaving 
his footprint. Bad news: if the footprint is detected, that happens 
interaction-free; the thief is not caught.} Of course, the Optical Theorem 
focuses just on that aspect of unitarity: any reaction on scattering, 
including absorption, must be accompanied by elastic forward scattering. The 
final state is a superposition of three orthogonal components, with their 
respective detectors, that of the third line being -- if you insist -- the 
bomb. Macroscopic quantum measurements are forced to choose one of 
the detectors.

One can also notice that there is nothing specific to an absorber in the 
above equations: the object can be a perfect mirror as well, sending the 
incoming photons to some outgoing mode \(c\) irreversibly leaving the 
interferometer and occasionally caught by some detector \(D_c\), as done
actually in Ref. \cite{kwhz} (as a matter of fact, ``outgoing" and 
``irreversible" can be used as  synonyms in the present case). Then line 
3 of the state vector is the amplitude along basis vector 
\(|0_a0_b1_c\rangle\), all the rest of the formulas remaining unchanged.

From the above discussion we see that with the bomb present, what appears 
at \(D_a\) is the forward-scattered wave, produced by the object-photon 
interaction on mode \(b\) and partially transferred to mode \(a\) by the 
second beamsplitter. If that component is detected, we have IFM in the 
now standard sense. 

For a perfect absorber, the forward-scattered wave is locked both in amplitude 
and phase, since it has to cancel the incoming wave, therefore the 
information conveyed by IFM is just about the presence or absence of the 
absorber (bomb): a black-and-white silhouette of it. In that particular 
limiting case the insertion of the absorber can be regarded as a modification 
of the boundary conditions obeyed by the wave field, limited by reflecting 
and absorbing diaphragms and walls anyway.

Interferometry, as a tool of detecting the presence of objects, is a technique 
of contrast enhancement; in the case discussed so far, that happens by means 
of producing a dark-field image on port \(D_a\) where the background due to 
the incoming wave has been extinguished. However, interferometric detection 
has better chances with at least partially transparent objects, where more 
amplitude contrast can be produced out of phase shifts. In that case, IFM 
would mean rejecting the phase information available in the statistics of 
\(D_a\) and \(D_b\) counts. 

Even if \(D_a\) and \(D_b\) are imperfect detectors, they are detectors of 
the same nature, and therefore can be calibrated together to the 
no-absorber case\footnote{That is not the case for neutrons, since mode 
\(b\), undergoing two more beamsplitter reflexions, is significantly damped 
with respect to mode \(a\)\cite{hs}.}. Then, using the statistics of 
repeated measurements, one can evaluate the respective calibrated 
detection probabilities \(P_1\) and \(P_2\).

The simplicity of the above three-line state vector is lost in the general 
case: besides absorption with a well-defined final state and transmission with 
some phase shift, all kinds of scattering events are possible. What can 
be determined from an extended IFM scheme is the total probability \(W\) 
of absorption and non-forward scattering:

\be
W = 2(1-P_1-P_2),
\label{IFMtot}
\ee

\n the factor 2 (to be modified for asymmetric beamsplitters) representing 
the presence of the object-free interferometric path. Besides the above, the 
detector counts contain information about the phase shift $\chi$ caused by 
the object to the transmitted mode-\(b\) photons: that can be extracted 
as

\be
\cos\chi = (P_2-P_1)/\sqrt{1-W}.
\label{phase}
\ee

The latter formulas are trivial adaptations of those used to evaluate the 
simplest case of interferometric IFM (not called IFM that time): the 
so-called ``stochastic absorption" of neutrons\cite{srt}. To make contact with 
the above formalism, let us introduce the object transmission amplitude 
$te^{i\chi}$; the sequence of events generated by the object sandwiched 
between two beamsplitters is now

\be
\pmatrix{1\cr 0\cr\vdots\cr} 
    \rightarrow\pmatrix{1/\sqrt2\cr i/\sqrt2\cr  \vdots\cr}
  \Rightarrow\pmatrix{\frac{1}{\sqrt2}\cr\frac{i}{\sqrt2}~t~e^{i\chi} \cr\vdots \cr}
    \rightarrow\pmatrix
	{\frac{1}{2}(1-t~e^{i\chi})\cr 
	 \frac{i}{2}(1+t~e^{i\chi})\cr 
	 \vdots\cr},
\label{rich}
\ee

\n where the dots represent all those channels - absorption, reaction or 
scattering modes, excluding forward-scattering - which get populated by the 
action of the object. Subtracting Equation (\ref{nobomb}) from 
Equation (\ref{rich}), the first two lines of the difference are the 
forward-scattered wave -- a colored silhouette of the object, carrying 
phase information as well -- which can be fully reconstructed from the 
interferometric measurement through formulas (\ref{IFMtot}) and (\ref{phase}). 
Let us observe that here, like in the perfect absorber case, the output on 
\(D_a\) is just forward-scattered photons processed by beamsplitter \(BS_2\), 
whereas \(D_b\) output is resulting from interference between the incoming 
and forward-scattered waves. 

Quantum silhouettes -- even if not called that way -- are much more 
widespread than that. In quantum jump phenomena, the unitary silhouette of 
a weak transition is observed in the fluorescence signal of a strong 
one\cite{deh}. In particle physics, unitarity can be used to draw inference 
about the contributions of non-detected processes\cite{part}. Last, not least: 
according to Mott's explanation of track formation in a cloud 
chamber\cite{mo}, individual ionization events are stringed into a track by 
their accompanying forward-scattered waves, taking the shape of a geometrical 
shadow -- a well-drawn silhouette -- for sufficiently high energies. 

In conclusion we note that interaction-free measurement is but an 
unusual combination of familiar features of quantum mechanics. Our description 
in terms of forward scattering follows in a way unavoidably from the 
principles of superposition and causality: nothing can extinguish an incoming 
wave except a forward scattered one, destructively interfering with 
it. The forward scattered wave is indeed there, being a by-product of 
absorption. 

The remaining ambiguity is of philosophical nature. State vector reduction -- 
or, if you prefer to call it that way, wave-particle duality -- makes 
absorption and any accompanying phenomena not-happened in those {\sl 
individual} cases when a non-absorbed particle has been detected. In the 
{\sl statistics} of manyobservations, the non-happened 
(``counterfactual"\cite{v}) events do appear through their unitary silhouette. 
From that point on, it is rather a matter of taste whether you consider 
unitary evolution a true physical process -- that view is implicite in the 
present paper -- or a bookkeeping for measurements. Anyway, the label ``interaction-free" is certainly correct in the technical sense of supporting 
the observation of fragile objects\cite{summ}.

\acknowledgments
I am indebted to Anton Zeilinger, Harald Weinfurter, Andor Frenkel, 
Johann Summhammer and P\'eter Domokos for enlightening discussions on 
the subject. I thank Andr\'as Patk\'os and Ferenc Csikor for discussing 
possible connections with particle physics. This work has been partially 
supported by the Hungarian Research Foundation (grant No. OTKA 
\break T 17312).

\end{document}